# Pronóstico de inflación de corto plazo en Argentina con modelos *Random Forest*

Federico D. Forte[1]
**Septiembre 2024**

## Abstract

El presente trabajo examina el desempeño de los modelos *Random Forest* para pronosticar la inflación mensual de corto plazo en Argentina, especialmente para el mes en curso o el inmediato posterior. Utilizando una base de datos con indicadores en frecuencia mensual desde 1962, se concluye que estos modelos alcanzan una precisión de pronóstico estadísticamente comparable al consenso de analistas de mercado relevado por el Banco Central de la República Argentina (BCRA) y a los modelos econométricos tradicionales. Una ventaja que ofrecen los Random Forest es que, al ser modelos no paramétricos, permiten explorar efectos no lineales en la capacidad predictiva de ciertas variables macroeconómicas sobre la inflación. Se encuentra, entre otras cosas, que: 1) la relevancia relativa de la brecha cambiaria para pronosticar la inflación crece cuando la brecha entre el tipo de cambio paralelo y el oficial supera 60%; 2) el poder predictivo del tipo de cambio sobre la inflación aumenta cuando las reservas internacionales netas del BCRA son negativas o cercanas a cero (específicamente, menores a USD 2.000 millones); 3) la relevancia relativa de la inflación rezagada y de la tasa de interés nominal para pronosticar la inflación del mes siguiente crece cuando aumenta el nivel de inflación y/o el nivel de la tasa de interés.

Palabras clave: Inflación, Random Forest, Pronóstico, Machine Learning, Econometría.

Clasificación JEL: C14, E31, E37.

---

1: Economista principal de BBVA Research, Banco BBVA Argentina. E-mail: federico.forte@bbva.com.

# Introducción

La inflación en Argentina ha experimentado una creciente aceleración desde principios de siglo, tras el fin del régimen de Convertibilidad. Particularmente, después de la pandemia la inflación pasó desde niveles en torno a 50% anual en 2021 a orillar 100% en 2022 y superar 200% en 2023, para luego rozar 300% en el primer semestre de 2024 (Gráfico 1), valores que no se experimentaban en el país desde las hiperinflaciones de 1989-90. La economía desde fines de 2022 muestra las características propias de un régimen de alta inflación (Frenkel, 1989): un contexto en el que el sistema económico se encuentra muy adaptado a la inflación, con elevada indexación de contratos y sustancial acortamiento de los plazos de repactación de precios y salarios, lo cual aumenta la propagación de shocks en los precios relativos. Fundamentalmente, ante la ausencia de anclas nominales alternativas, los precios de la economía tienden a coordinarse en buena medida ante los movimientos en el tipo de cambio.

En este contexto de acelerada volatilidad de precios, se ha vuelto crecientemente relevante para la toma de decisiones presupuestarias de las empresas y de los agentes económicos la predicción de los valores mensuales (e incluso semanales) de inflación en alta frecuencia y "en tiempo real". La inflación mensual pasó de promediar 3% entre 2018 y 2021, a promediar más de 6% entre 2022 y mediados de 2023, para luego saltar a niveles de dos dígitos mensuales desde el segundo semestre de 2023 hasta el primer trimestre de 2024 (Gráfico 2).

En este entorno, técnicas econométricas que en contextos de baja inflación exhiben errores de pronóstico de, como mucho, unas décimas en la medición mensual, fácilmente pueden incurrir en errores de varios puntos porcentuales por mes. Esos errores pueden implicar en la práctica severas pérdidas o ganancias para empresas que necesitan tomar decisiones de remarcación de precios, o (como es por ejemplo el caso de las entidades bancarias) acarrear sustanciales ganancias o pérdidas contables debido al ajuste por inflación de los balances.

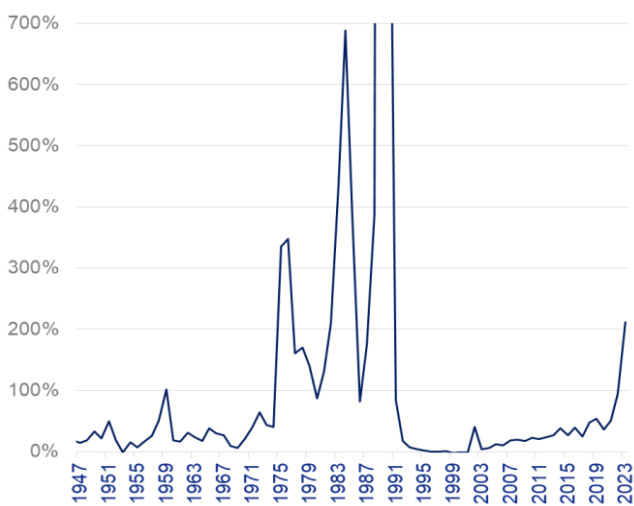

Gráfico 1. **INFLACIÓN EN ARGENTINA 1947-2023** (VAR. % ANUAL)

Fuente: INDEC. Para el período 2007-15 se utilizó un promedio de indicadores provinciales.

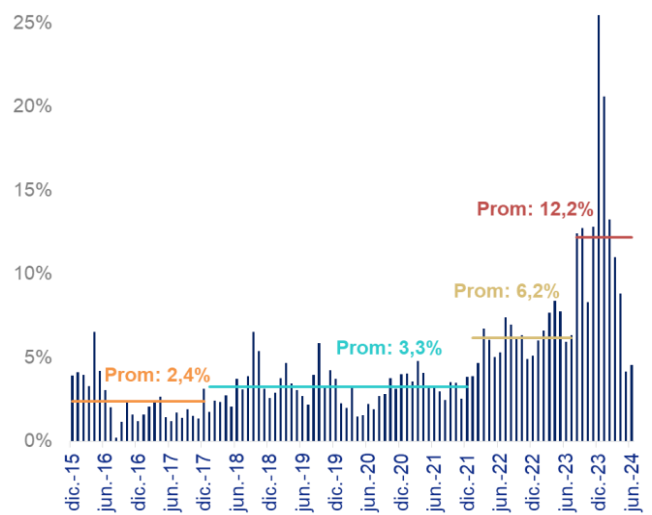

Gráfico 2. **INFLACIÓN MENSUAL** (VAR. % M/M)

Fuente: INDEC.

Por ese motivo, en este trabajo se propone explorar modelos de *Random Forest*, una técnica muy difundida de aprendizaje automático (*Machine Learning*)[2], para evaluar su desempeño en el pronóstico de corto plazo de la inflación en Argentina (i.e., el mes en curso o el inmediato posterior) y extraer conclusiones respecto de su aplicabilidad y valor añadido que podrían ofrecer como herramienta complementaria dentro del conjunto de metodologías disponibles para predecir la evolución de los precios en un entorno tan complejo y desafiante. En la sección siguiente se describen las principales características de la metodología elegida, y a continuación se detallan los datos utilizados. Luego, se resumen los principales resultados obtenidos, junto con especificaciones técnicas

---

2: A lo largo de este documento, los términos "aprendizaje automático" y "Machine Learning" se utilizarán de manera intercambiable.

(por ejemplo, los hiperparámetros) del modelo aquí estimado. Finalmente, se plantean algunas conclusiones, limitaciones y líneas de investigación futuras.

## Metodología

El modelo Random Forest (Breiman, 2001) es una técnica de aprendizaje automático muy utilizada en la actualidad, tanto para resolver problemas de clasificación de datos (entre distintas clases y/o grupos, donde la variable dependiente está definida en números discretos) como para realizar ejercicios típicos de regresión, del estilo que tradicionalmente han sido abordados mediante técnicas econométricas clásicas (como Mínimos Cuadrados Ordinarios -MCO-, por ejemplo). Estos modelos han ganado mucha popularidad especialmente cuando se trabaja con grandes bases de datos de muchas variables y dimensiones (a lo que hoy en día se le suele referir comúnmente como *big data*), porque permiten analizar relaciones complejas no lineales y requieren poca intervención del investigador (imponiendo o asumiendo relaciones estructurales previamente) para obtener resultados robustos (Goulet Coulombe *et al*, 2022), entre otras ventajas que detallaremos más adelante.

En este trabajo, la "variable objetivo" (*target*) que se busca pronosticar es la inflación para el mes en curso (o a un mes vista como máximo), $\pi_t$, estimando la potencial relación no lineal con sus rezagos y un conjunto de regresores, $X_{t-h}$ (que pueden ser contemporáneos o rezagados):

$$\pi_t = f(\pi_{t-1} \ldots \pi_{t-h}; X_t \ldots X_{t-h}) + e_t$$

Donde $e_t$ es un término de error y $f(.)$ una función no lineal sin forma paramétrica asumida *a priori*. En base a esta estimación luego se obtiene un pronóstico para $\pi_t$:

$$\pi_t = f(\pi_{t-1} \ldots \pi_{t-h}; X_t \ldots X_{t-h})$$

Es importante destacar desde el principio que el propósito de estas estimaciones no es bajo ningún punto de vista identificar relaciones causales entre las variables, sino que toda la tarea está orientada hacia fines meramente de pronóstico, por lo que problemas de simultaneidad entre las variables no implican un obstáculo para su inclusión dentro del modelo.

La estimación de los modelos Random Forest se basa en la construcción y combinación de múltiples "árboles de decisión". Cada árbol de decisión se estima subdividiendo la muestra de la variable objetivo ($\pi_t$) de acuerdo a distintos umbrales dentro de los dominios de las variables regresoras, de forma tal de obtener la mejor precisión posible para predecir $\pi_t$ en cada paso ante cada subdivisión. El procedimiento sigue a grandes rasgos la siguiente mecánica (Gráfico 3):

1. Partiendo de un "nodo raíz" $h_0$, en el que se encuentra la muestra entera de $\pi_t$, se asume la existencia de un vector de regresores que aquí llamaremos $Z = (\pi_{t-1} \ldots \pi_{t-h}; X_t \ldots X_{t-h})$, con $d$ dimensiones (es decir, hay $d$ variables regresoras en $Z$ para predecir la variable objetivo) y N observaciones para cada una.

2. Se evalúa qué partición posible de las observaciones dentro de $\pi_t$ minimiza la diferencia entre la mediana de cada partición y el subconjunto de observaciones de $\pi_t$ que quedaron dentro de cada partición. Para hacer esto, se ejecuta el siguiente procedimiento para cada valor posible de cada una de las $j_i = j_1 \ldots j_d$ dimensiones que existen dentro del vector $Z$:

    2.1. Se ordenan las observaciones de $\pi_t$ de acuerdo a los valores en la variable $j_i$.
    2.2. Subdividir las observaciones en dos subconjuntos. Hay como mucho N-1 umbrales posibles para hacer esta división en dos partes de la muestra que llegó al nodo $h_0$.
    2.3. Dada una división de la muestra en dos "nodos hojas" $h_1$ y $h_2$ se computa la mediana de la variable objetivo en cada subconjunto de datos y el Error Absoluto Medio (MAE)[3] entre dichas medianas y los datos observados para la variable target en cada $h_1$ y $h_2$.
    2.4. Se suman ambos MAE, obteniendo así una medida del error de pronóstico del árbol derivado de esa subdivisión en particular.

---
3: Se pueden usar otras medidas de error (como el error cuadrático medio o el R2), pero seleccionamos el MAE por ser el más robusto ante valores atípicos.

2.5. Se elige el mejor umbral de los (N-1) x $d$ test probados en forma iterativa, en función del que obtenga el menor MAE. Así se obtiene una primera subdivisión del "nodo raíz" en dos "nodos hoja".

3. Se repite este procedimiento para cada nodo hoja, mejorando así el MAE del árbol total en cada paso.

4. Se establecen criterios de parada para este procedimiento, como una profundidad máxima del árbol, un número mínimo de datos por hoja o un umbral mínimo de reducción en el MAE. En este trabajo establecimos un límite máximo para la profundidad del árbol.

Los árboles de regresión por sí solos tienden a tener problemas de sobreajuste (es decir, se ajustan tanto a los datos de la muestra que tienden a tener un desempeño pobre para predicción fuera de muestra). Por este motivo, los modelos random forest buscan solucionar este problema estimando múltiples árboles, y luego la estimación final se calcula ya sea por "votación" entre todos los árboles en caso de problemas de clasificación, o tomando el promedio o mediana de los resultados de todos los árboles en el caso de regresión. Esta técnica combinada permite entonces mayor robustez y menor sobreajuste a la muestra.

Gráfico 3. **ESQUEMA: EJEMPLO DE CONSTRUCCIÓN DE UN ÁRBOL DE DECISIÓN**

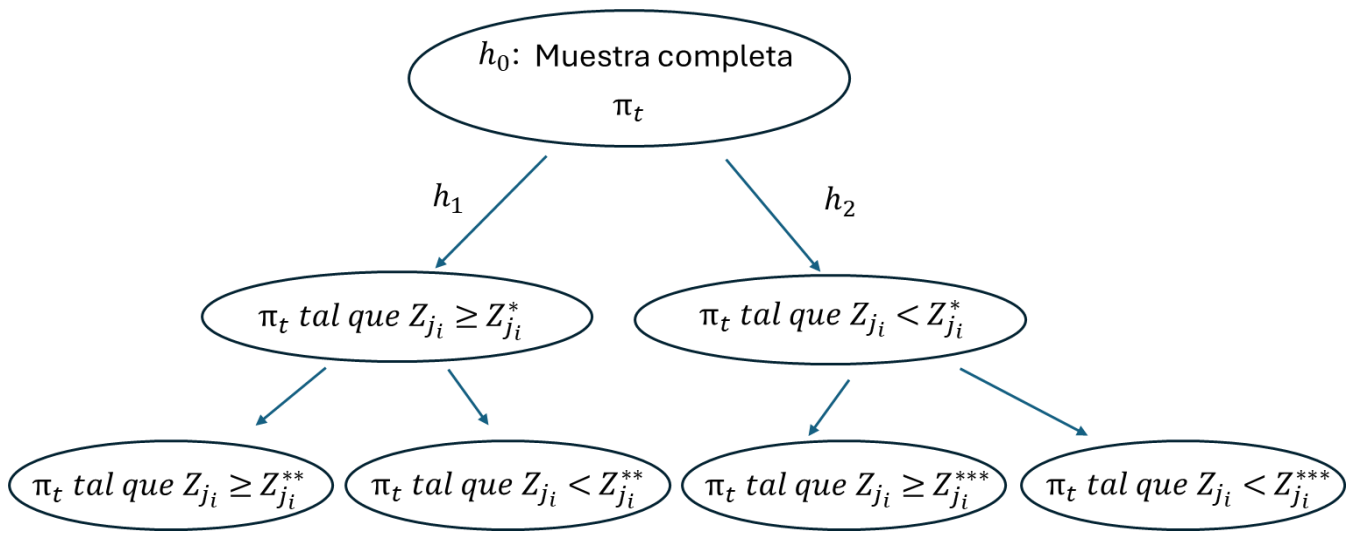

Fuente: Elaboración propia.

Asimismo, otra estrategia para limitar el sobreajuste consiste en imponer un límite a la cantidad de variables que se prueban en los tests de definición de umbral en cada hoja. En lugar de evaluar umbrales para todo $d$ en cada hoja, aquí se optó por testear el 30% de las variables, seleccionadas en forma aleatoria, en cada hoja de cada árbol (lo cual asimismo reduce la demanda de procesamiento computacional para el modelo).

De esta explicación se desprende que la metodología detrás de los modelos *random forest* permite estimar entonces "comportamientos promedio" de la variable objetivo $\pi_t$ ante distintos umbrales de cada variable incluida dentro del vector $Z$. Esto implica la posibilidad de analizar relaciones no lineales entre las variables, sin la necesidad de imponer y/o asumir estructuras específicas de antemano por parte del investigador.

A la vez, los modelos Random Forest permiten la incorporación de una multiplicidad de indicadores, lidiando adecuadamente con potenciales problemas de sobreajuste, presencia de valores atípicos (*outliers*) y datos faltantes (Biau y Scornet, 2016; Hastie et al., 2009). Son modelos no paramétricos que han demostrado ser robustos ante varios problemas estadísticos típicos como la multicolinealidad entre regresores y la heterocedasticidad, que suelen afectar crucialmente el desempeño de los modelos paramétricos (como las técnicas econométricas tradicionales). Tal como otros métodos de Machine Learning, los modelos Random Forest pueden procesar un gran número de variables y se ven poco afectados por la inclusión de variables no significativas (Hastie et al., 2009). Adicionalmente, la definición de sus hiperparámetros no demanda demasiada complejidad computacional o muestras extremadamente grandes, como sí es necesario en el caso de las redes neuronales, mientras que muestra en muchas ocasiones un rendimiento comparable en términos predictivos a las redes neuronales o el *boosting generalizado* (Jones et al., 2017).

Los modelos Random Forest, a pesar de ser ampliamente populares hace ya más de una década en múltiples disciplinas (por ejemplo, ecología -Cutler et al., 2007-, medicina -Klassen et al, 2008-, agricultura -Löw et al, 2012- y transporte -Zaklouta et al, 2011-), su incorporación para pronóstico de variables macroeconómicas es más reciente (Gawthorpe, 2021; Yoon, 2021; Goulet Coulombe et al, 2022). Entre los desarrollos más modernos está el trabajo de Goulet Coulombe (2024) que va un paso más allá y desarrolla un "Random Forest Macroeconómico", adaptando este algoritmo canónico a un modelo de parámetros móviles generalizados no sólo para pronosticar variables macroeconómicas sino que también busca comenzar a aplicar este herramental para intentar captar parámetros estructurales de modelos macroeconómicos.

El trabajo de Lenza et al (2023) aplica modelos random forest por cuantiles para pronosticar la inflación de la Eurozona a un mes vista (no sólo producen estimaciones puntuales, sino que que estiman toda la función de densidad de dichos pronósticos), concluyendo que con este método se alcanzan pronósticos de precisión competitiva con los mejores modelos econométricos más tradicionales y a la vez estas estructuras permiten examinar no linealidades entre la inflación y los regresores incluidos.

A partir de estos desarrollos y conclusiones recientes de la literatura, este trabajo tiene entonces el propósito de estimar un modelo de Random Forest para pronosticar la inflación de corto plazo en Argentina (1 mes adelante como máximo), buscando alcanzar una precisión que sea competitiva con otros modelos econométricos tradicionales (como los modelos autorregresivos de medias móviles -ARMA-), y que a la vez requiera mucha menor necesidad de calibración o estructura teórica que por ejemplo los modelos dinámicos y estocásticos de equilibrio general (DSGEs). Asimismo, la utilización de esta metodología permitirá examinar no linealidades que son más difíciles de captar con otros tipos de metodologías.

## Datos

La base de datos utilizada para este trabajo se compone de trece variables económicas detalladas en la Tabla 1, todas medidas en frecuencia mensual y desde el año 1962. La variable objetivo es la inflación mensual nacional publicada por el organismo oficial de estadísticas de Argentina (INDEC). Las variables predictoras (vector $Z$) se seleccionaron tomando como inspiración el típico marco teórico de la curva de Phillips y se incluyeron medidas de presión de costos e inflación internacional. Específicamente, se incluyeron indicadores de las condiciones monetarias de la economía (tasas de interés y cantidad de dinero), actividad económica, salarios, variables asociadas al mercado cambiario (tipo de cambio oficial, paralelo, reservas internacionales netas) y precios internacionales (inflación de EE. UU., junto con los precios del petróleo y el trigo).

Como la inflación en Argentina desde el año 2022 alcanzó niveles que no se veían desde la década de 1980, fue necesaria para mejorar la precisión del modelo *random forest* la inclusión de variables con información disponible desde antes de la Convertibilidad (que fue implementada en abril de 1991). Sólo de este modo el modelo pudo incorporar y predecir dinámicas y comportamientos propios de entornos de alta inflación e hiperinflación. Si solo se incluían datos posteriores a 1991, las inflaciones mensuales entre mediados 2022 y principios de 2024 estaban por fuera del dominio de la variable objetivo y por ende no existía la posibilidad de pronosticar correctamente esos valores con esta metodología en particular.

La necesidad de incorporar una historia pasada tan larga en frecuencia mensual reduce drásticamente los indicadores disponibles relativamente homogéneos. Por esta falta de información no se han incluido variables que midan expectativas u otras aristas posibles del entorno económico. De todas formas, la inclusión del tipo de cambio paralelo y la inflación rezagada como predictores podrían tener cierta correlación con las expectativas de inflación en la economía.

La construcción de la mayoría de las variables es fácilmente reproducible en base a la información disponible en la Tabla 1. El único indicador que requirió una estimación propia de cierta complejidad fue la actividad económica en frecuencia mensual antes de 1993. Para estimar ese período, se utilizó el componente principal que explica la mayor proporción de varianza entre la evolución de la producción de acero, automotriz y crédito total en pesos medido en términos reales. La evolución de este indicador presenta una correlación superior al 80% con el Estimador Mensual de la Actividad Económica de INDEC desde 1993 en adelante.

Tabla 1. **BASE DE DATOS: INDICADORES SELECCIONADOS**

| | Variable | Unidad de medida | Fuente / Detalle |
|---|---|---|---|
| Inflación (target) | IPC Nacional Argentina | Variación mensual | Índice general de INDEC. Entre 2007 y 2015: promedio indicadores provinciales, Congreso, CABA. |
| Actividad económica y salarios | Salarios registrados | Variación mensual | Índice de Salarios Básicos de la Industria y la Construcción (ISBIC) desde 1962. Remuneración imponible promedio de los trabajadores estables (RIPTE) desde 1994. |
| | Actividad económica | Variación mensual (promedio móvil 3 meses) | Estimador Mensual de la Actividad Económica (EMAE) de INDEC (desde 1993) y proxy mensual en base al crédito real, producción automotriz y acero desde 1962. |
| Política monetaria | Tasa de interés | Tasa efectiva mensual | Tasa de plazo fijo hasta 60 días entre 1962 y 2015. Tasa de política monetaria relevante desde 2015. Promedios mensuales. |
| | Base monetaria | Variación mensual | BCRA |
| | Agregado M2 Total | Variación mensual | BCRA |
| Precios internacionales | Trigo | Variación mensual | Precio internacional de la tonelada trigo (Haver) |
| | Petróleo | Variación mensual | Precio internacional del barril, tipo Brent (Haver) |
| | Inflación EE. UU. | Variación mensual | Bureau of Labor Statistics |
| Mercado cambiario | Tipo de cambio Oficial | Variación mensual | BCRA |
| | Tipo de cambio Paralelo | Variación mensual | Tipo de cambio paralelo relevante en cada período (Alphacast, FIEL, Ámbito Financiero). |
| | Brecha cambiaria | Diferencia entre el tipo de cambio paralelo y el oficial (en %) | - |
| | Reservas internacionales netas | Millones de dólares a precios constantes | Stock de reservas brutas menos pasivos de corto plazo (12 meses vista) en moneda extranjera del BCRA. Estimación BBVA Research. |

Fuente: Elaboración propia.

# Resultados y especificaciones del modelo

Para la estimación de los pronósticos a partir del modelo Random Forest, se optimizó primero el valor de algunos hiperparámetros clave. Para ello se aplicó el procedimiento estándar de subdividir aleatoriamente la base de datos en un conjunto de entrenamiento y un conjunto de test (80% y 20% del total de los datos respectivamente). Los hiperparámetros del modelo se optimizaron en base a los datos del conjunto de entrenamiento minimizando el MAE[4]. Se concluyó que con 500 árboles por random forest y 15 hojas para la profundidad máxima de cada árbol se obtenían los mejores resultados en términos de MAE.

Aún fijando estos hiperparámetros con dicho procedimiento, existe cierta volatilidad en los pronósticos resultantes dependiendo del conjunto de entrenamiento puntual (80% de la base) que queda seleccionado aleatoriamente en cada estimación[5]. Por ello, para el cálculo de cada pronóstico mensual se repite 25 veces este procedimiento, re-muestreando el conjunto de entrenamiento en cada iteración, y se obtiene así un promedio y un desvío estándar de los pronósticos para cada mes. En cada iteración se evalúa también la precisión del modelo para pronosticar el

---

4: Se aplicó el procedimiento de validación cruzada (*cross-validation*) para tornar más robusta la estimación de los hiperparámetros. Consiste en dividir el conjunto de entrenamiento en N subconjuntos disjuntos elegidos al azar (en este caso se dividió en 10), y usar cada uno de ellos como conjunto de validación de los modelos entrenados en base a la suma de los N-1 subconjuntos restantes. La precisión promedio de ese ejercicio iterado 10 veces es lo que se tomó como referencia para evaluar la precisión del uso de distintos hiperparámetros.
5: El modelo se re-estima siempre en base al 80% de los datos disponibles en cada momento del tiempo. Para evitar problemas de sobreajuste no se usa un 20% de los datos que quedan incluidos aleatoriamente en el conjunto de test.

conjunto de test (20% remanente de los datos, que se usa como estrategia "*pseudo-out-of-sample*" de evaluación de capacidad de pronóstico), midiéndose también esta a través del MAE alcanzado al predecir los datos en test.

La metodología del Random Forest nos permite estimar la importancia relativa de cada variable para mejorar la precisión del pronóstico, en base del promedio de esa mejora para todos los árboles al computar sus respectivas hojas (Gráfico 4). De aquí se concluye que según este modelo los factores asociados a lo que comúnmente se conoce como "inercia inflacionaria" (medida como la inflación rezagada uno y/o dos períodos) son de los más relevantes para mejorar el pronóstico de la inflación mensual. En segundo lugar, los factores asociados a las condiciones monetarias de la economía (la tasa de interés y la cantidad de dinero), y, luego, los referidos a los salarios, al mercado cambiario y la actividad económica. Los menos relevantes para aumentar la precisión del pronóstico son los precios internacionales. Cabe aclarar una vez más que estas estimaciones no implican bajo ningún punto de vista una noción de causalidad, sino que simplemente su desempeño se evalúa a los fines de pronóstico.

Gráfico 4. **IMPORTANCIA RELATIVA DE LAS VARIABLES PREDICTORAS PARA MEJORAR EL PRONÓSTICO DE INFLACIÓN EN EL MOMENTO "T"** (REDUCCIÓN PROMEDIO DEL ERROR ABSOLUTO MEDIO)

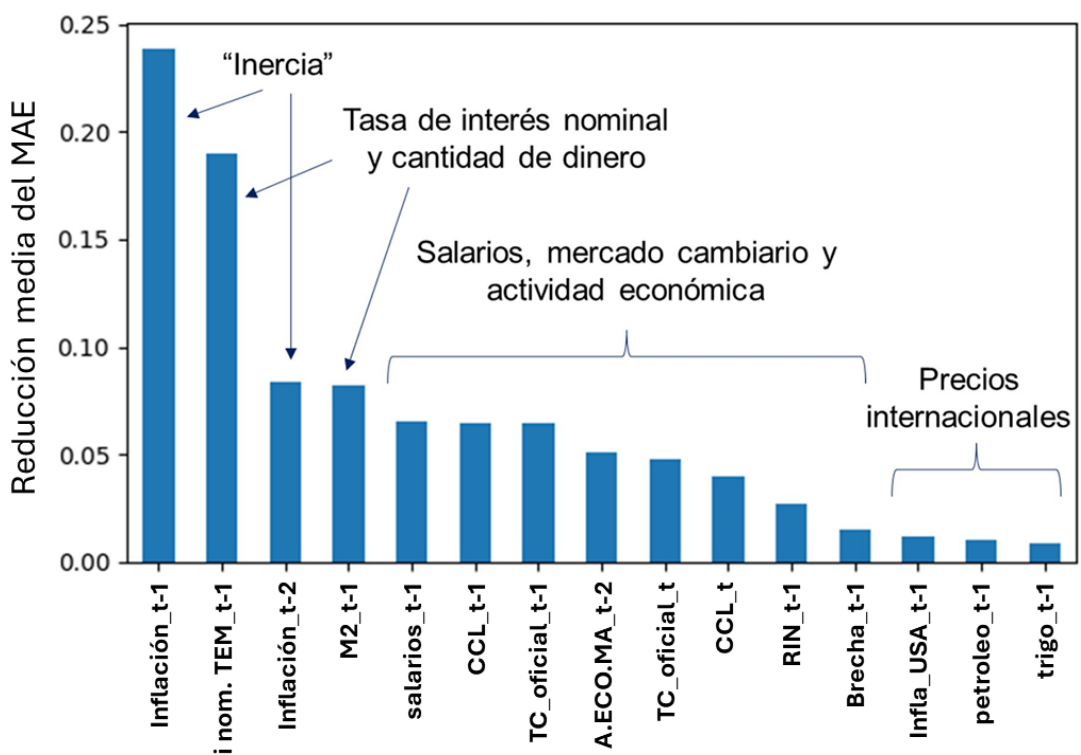

Definición de las variables: "Inflación_t-1" y "Inflación_t-2" se refieren a la inflación mensual rezagada 1 y 2 meses respectivamente; "i nom. TEM" es la tasa de interés nominal efectiva mensual; "M2" es el agregado monetario M2 total; "salarios" refiere a los salarios nominales; "CCL" es el tipo de cambio paralelo; "TC_oficial" es el tipo de cambio del mercado oficial de Argentina o en su defecto el aplicable a comercio exterior; "A.ECO.MA" es la actividad económica, medida en promedio móvil 3 meses de la variación mensual; "RIN" son las reservas internacionales netas del BCRA medidas a precios constantes en dólares; "Brecha" es la brecha entre el tipo de cambio oficial y el paralelo; "infla_USA" es la inflación mensual de EE.UU.; "petróleo" y "trigo" refieren a la variación mensual de sus respectivos precios internacionales. Cuando aparece la leyenda "t" implica que la variable se incluye en forma contemporánea a la inflación objetivo, mientras que "t-1" o "t-2" implican que las variables están incorporadas con rezago de 1 o 2 meses respectivamente. La unidad de medida de cada variable se explica en la Tabla 1.
Fuente: Elaboración propia.

Para la selección de estas variables se evaluó el desempeño del modelo de acuerdo a su precisión para pronosticar las observaciones del conjunto de test (minimizando el MAE), incluyendo distintas combinaciones de rezagos y reemplazando variables altamente correlacionadas entre sí (por ejemplo, M2 y Base Monetaria). Con la inclusión de la tasa de interés nominal junto con la inflación rezagada como regresores se logró mejor precisión que incorporando distintas variaciones y/o rezagos de diferentes medidas directas de tasa de interés real.

Con el objetivo de comparar los pronósticos resultantes de este modelo con otras alternativas y metodologías, se computó el pronóstico a 1 mes vista *out-of-sample* (i.e., suponiendo que el mes pronosticado está fuera de la muestra que se utiliza para la estimación del modelo en cada mes), para los últimos 24 meses de la base de datos (el período desde agosto 2022 a julio 2024), y en base a ello se compara el Error Absoluto Medio (MAE), tanto para el Random

Forest como para otros competidores. También se computó el promedio del Error Absoluto Medio en el conjunto de test derivado de cada una de las 24 estimaciones mensuales (Tabla 2)[6].

Tabla 2. **EVALUACIÓN DE LA PRECISIÓN COMPARADA DE DISTINTOS MODELOS Y DEL CONSENSO DE ANALISTAS**

|  | **Random Forest** | **ARMA** | **Lasso** | **Ridge** | **REM-BCRA** |
|---|---|---|---|---|---|
| Conjunto de test | 2,13% (0,50%) | N/A | 2,26% (0,37%) | 2,22% (0,44%) | N/A |
| Pronóstico a 1 mes (fuera de muestra) | 1,31% (0,42%) | 1,70% (0,31%) | 1,42% (0,47%) | 1,25% (0,48%) | 0,85% |

Error absoluto medio en el conjunto de test y del pronóstico a 1 mes vista fuera de muestra, entre ago-22 y jul-24; desvíos estándar entre paréntesis.
Fuente: Elaboración propia.

Se evaluó el desempeño de los pronósticos del modelo Random Forest versus el de modelos tradicionales ARMA y también con variantes muy utilizadas en la literatura de Machine Learning como son los modelos Lasso y Ridge[7]. Para la evaluación del modelo ARMA se estimó para cada mes el mejor modelo en base a los datos disponibles a cada fecha, optimizando la selección de rezagos y promedios móviles con el indicador de bondad de ajuste de Akaike, y se estimó el pronóstico a un mes vista.

Para el caso de Lasso y Ridge, se utilizó una metodología análoga a la del Random Forest: para cada mes se estimaron los hiperparámetros correspondientes que mejoraban la precisión del modelo (con *10-fold cross-validation*), y luego se calcularon los pronósticos mensuales re-muestreando 25 veces el conjunto de entrenamiento y test en cada mes, por lo que el pronóstico puntual mensual resulta ser el promedio de esas 25 iteraciones. Esto se hace porque la estimación se calcula utilizando un conjunto de entrenamiento de 80% de los datos (para evitar sobreajuste), y de esta manera se evita el posible riesgo de que el pronóstico sea muy sensible a la selección aleatoria de ese conjunto de entrenamiento en cada mes.

En la Tabla 2 se puede apreciar que el Random Forest muestra un desempeño muy similar al de sus competidores. Es ligeramente más preciso en promedio para predecir los conjuntos de test, dado que muestra el menor Error Absoluto Medio, pero los resultados para Ridge y Lasso se encuentran dentro del intervalo de 1 desvío estándar, lo que quiere decir que estadísticamente muestran un desempeño de pronóstico muy similar y no se puede determinar que uno sea sistemáticamente mejor que los otros en términos de predicción en test.

Al evaluar el desempeño de los pronósticos fuera de muestra a 1 mes vista, se observa que el Random Forest muestra un menor MAE (mejor performance de pronóstico) que los modelos tradicionales ARMA, con una diferencia de prácticamente 1 desvío estándar. Asimismo, su MAE alcanzado es marginalmente mejor que el modelo Lasso y ligeramente peor al modelo Ridge, pero como estos resultados no presentan una diferencia siquiera superior a un desvío estándar entre sí, no hay evidencia suficiente para afirmar que uno de estos modelos sea estrictamente superior para pronosticar la inflación de este período.

Complementariamente, comparamos el desempeño de pronóstico del Random Forest con la mediana de los pronósticos de analistas profesionales relevados en el REM (Relevamiento de Expectativas de Mercado) del BCRA, considerando siempre el pronóstico más cercano temporalmente al dato observado cada mes. De allí se desprende que el error absoluto medio de los pronosticadores ha sido 0,9% para la inflación mensual de la ventana temporal analizada, mientras que el Random Forest ha tenido un error medio de 1,3% +/- 0,4 p.p. A pesar que el consenso de analistas ha mostrado un desempeño promedio ligeramente mejor, la diferencia entre el modelo y la mediana de los pronosticadores es estadísticamente no significativa.

---

6: Se eligió este período para evaluar los pronósticos porque el objetivo es lograr la mayor precisión posible en el entorno particular de alta inflación que comenzó a presentarse en Argentina desde mediados de 2022.
7: Para la estimación de pronósticos con Lasso, Ridge y ARMA, se usaron las mismas variables que en el caso del Random Forest, con la única excepción del stock de reservas internacionales netas, que fue excluido por ser la única variable no estacionaria entre los indicadores incorporados.

Gráfico 5. **DEPENDENCIA PARCIAL DE LA INFLACIÓN MENSUAL RESPECTO DE LOS REGRESORES**

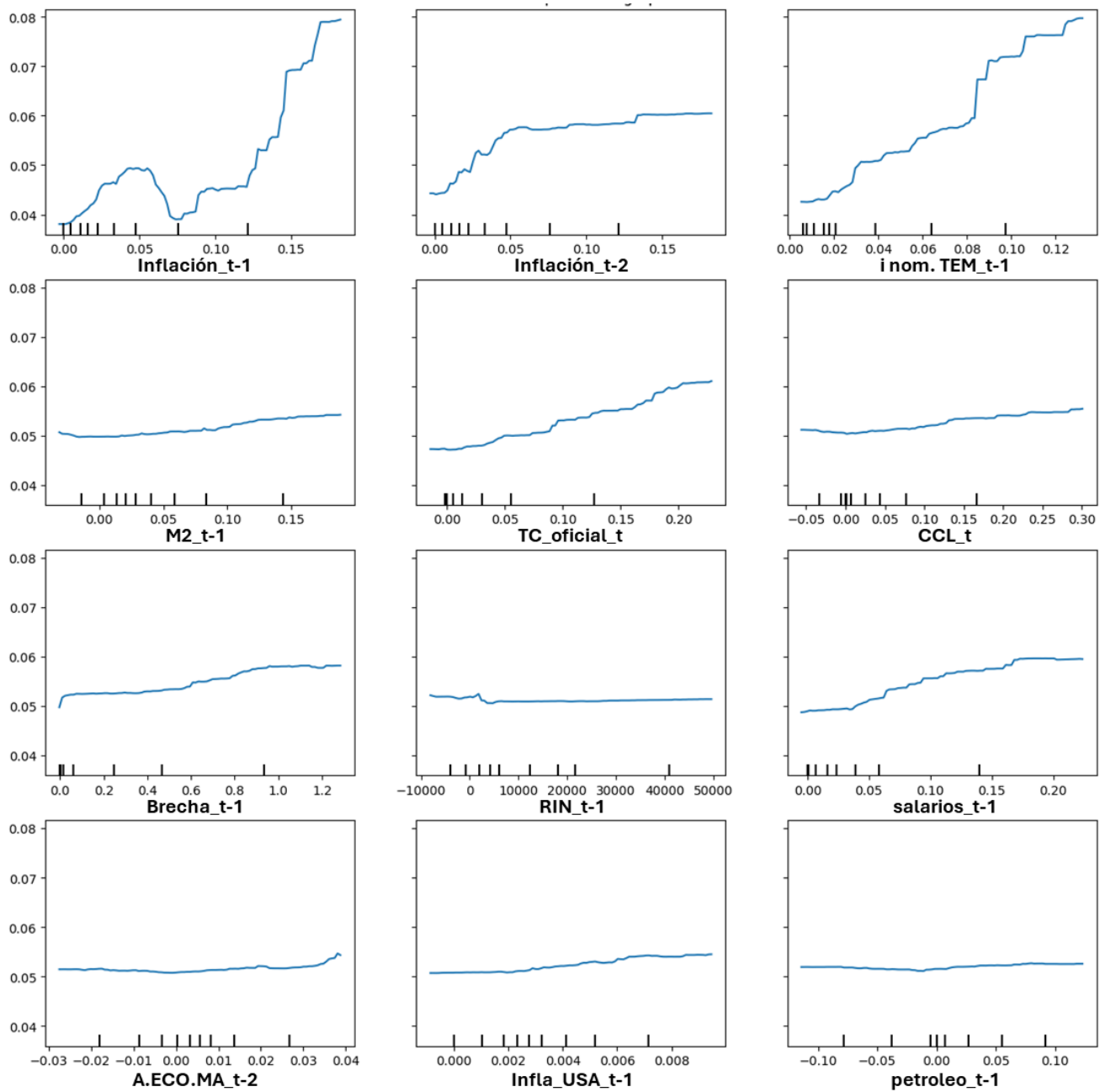

Fuente: Elaboración propia.

Una ventaja adicional de los modelos Random Forest es que permiten el análisis de no-linealidades en las relaciones entre la variable objetivo y los regresores. Para ello se computaron las dependencias parciales entre la inflación mensual y cada una de las distintas variables incorporadas en este modelo (Gráfico 5). De esta manera se estima la dependencia entre la respuesta de la variable target (en este caso, la inflación mensual) y el regresor de interés manteniendo constante el resto de las variables. Intuitivamente, puede ser entendida como la respuesta esperada del target en función del nivel en que se ubique cada regresor en forma aislada (muy semejante al concepto de "derivada parcial" pero trasladado al entorno de los modelos Random Forest).

De este modo se pudieron computar varios efectos no lineales que son complicados de captar con otras metodologías. En primer lugar, se ve que la inflación del mes inmediato previo (que en muchos estudios se usa comúnmente para medir el efecto de la "inercia inflacionaria" -por ejemplo, en Frenkel y Friedheim, 2017-) es crecientemente relevante para predecir la inflación del mes subsiguiente a medida que aumenta el nivel de inflación general. La inflación rezagada dos meses también crece en relevancia a medida que la inflación acelera a niveles más altos, pero luego se estanca su impacto marginal, lo cual se encuentra en línea con el hecho de que la inflación más alta es más volátil (hecho que se verifica empíricamente en múltiples estudios previos, por ejemplo, en Arndt y Enders, 2024), por lo que pierde correlación a partir de los dos meses de rezago.

Asimismo, la tasa de interés nominal es de las que mayor poder predictivo poseen, y en forma creciente a medida que la tasa crece, en línea con la clásica paridad de Fischer (Cochrane, 2016). Este resultado concuerda también con lo afirmado, por ejemplo, en Cagan (1956) y Heymann y Leijonhufvud (1995), acerca de que en entornos inflacionarios más elevados, las variables nominales tienden a correlacionarse más entre sí, a la vez que correlacionan menos con los movimientos de las variables reales.

Gráfico 6. **DEPENDENCIA PARCIAL DE LA INFLACIÓN RESPECTO DE DISTINTAS COMBINACIONES DE REGRESORES**

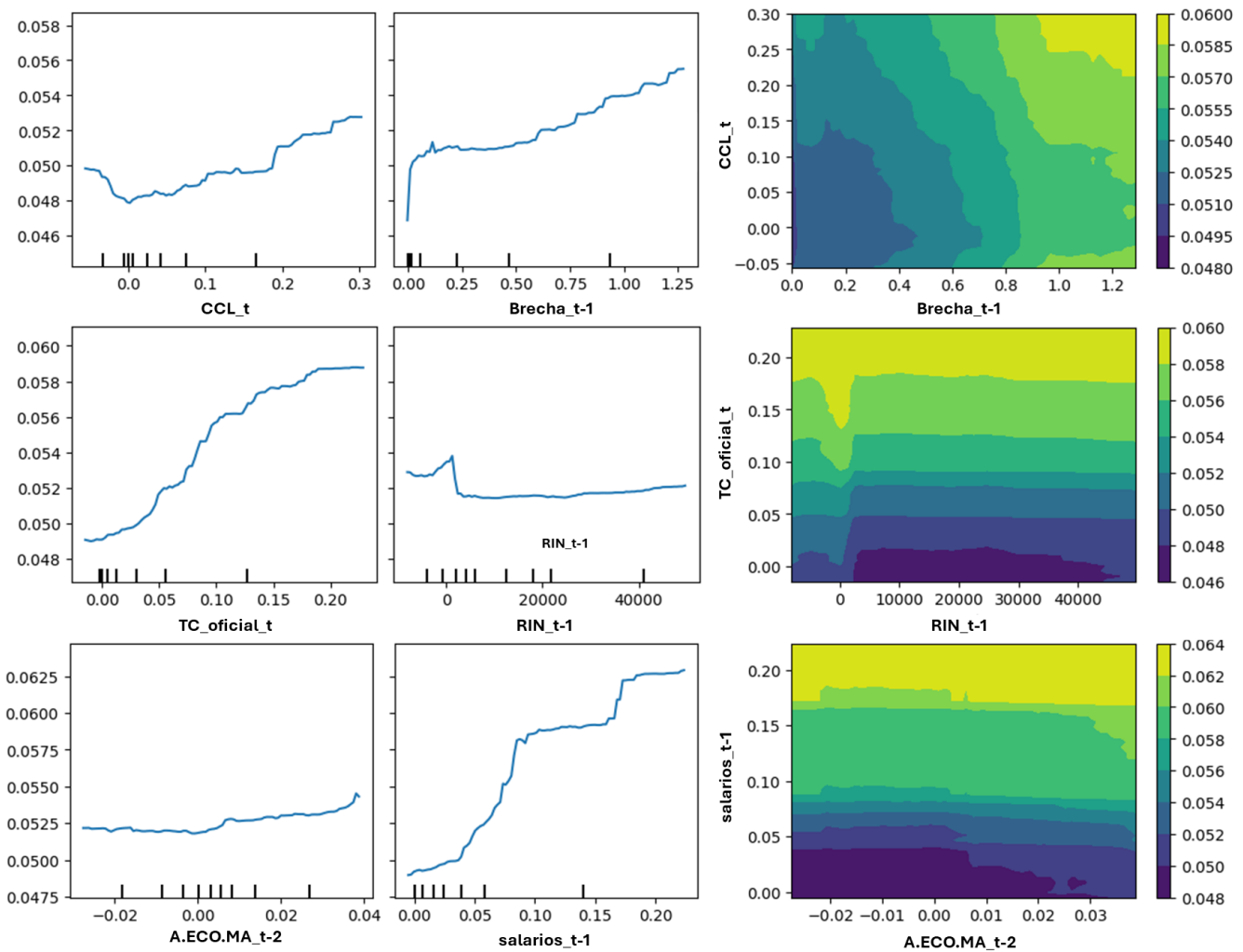

Fuente: Elaboración propia.

Luego, la respuesta de la inflación ante variaciones del tipo de cambio oficial es también mayor cuanto más alta es la variación del tipo de cambio en cuestión. Es decir, esto podría considerarse como evidencia adicional respecto de que el *pass-through* del tipo de cambio a precios es no lineal y creciente con el tamaño de la devaluación. Algo semejante ocurre en el caso de los salarios. En menor medida, también el tipo de cambio paralelo tiene un impacto creciente a medida que aumenta su variación, pero mucho más acotado en sus niveles absolutos que el efecto del tipo de cambio oficial. En este respecto, la brecha cambiaria también tiene mayor poder predictivo sobre la inflación a medida que aumenta de tamaño. Puntualmente, el umbral de 60% parece ser un punto de quiebre en el que el impacto aumenta en forma más significativa.

Asimismo, la respuesta de la inflación ante la variación de los salarios nominales también crece a medida que sus variaciones mensuales aumentan, fundamentalmente cuando las variaciones son superiores al intervalo entre 4 y 5% mensual. Por otra parte, el poder predictivo de la actividad económica y los precios internacionales no muestra no-linealidades sustanciales.

Gracias a esta metodología también se pueden visualizar no-linealidades combinadas entre varios regresores (Gráfico 6). A partir de allí se encuentra que el efecto de la variación CCL sobre la inflación se potencia junto con el de la brecha cambiaria a medida que crecen ambas variables.

Una de las no-linealidades combinadas más marcadas es la del tipo de cambio oficial y el stock de reservas internacionales netas. El poder predictivo de ambas variables sobre la inflación crece cuando las reservas netas son cercanas a cero o negativas. Es decir, a igualdad de tamaño de la devaluación del tipo de cambio, el poder predictivo de este sobre la inflación será mayor cuando las reservas netas se encuentren en el entorno de cero o son negativas. El umbral de reservas netas a partir del cual cambia cualitativamente la respuesta de la inflación se ubicaría en torno a los USD 2.000 millones, a valores constantes de julio 2024.

Finalmente, se encuentra que la respuesta de la inflación ante cambios en los salarios nominales es mayor cuando la actividad económica crece por encima de 0,8%. Se produce otro salto de escalón en el impacto cuando el aumento de la actividad supera el 2% mensual. A la inversa, cuando la actividad cae, el poder predictivo del aumento de salarios sobre la inflación disminuye.

# Reflexiones finales

El objetivo de este trabajo consistió en explorar el desempeño de los modelos Random Forest para pronosticar la inflación mensual de corto plazo en Argentina, fundamentalmente en el contexto del régimen de alta inflación que experimenta el país desde 2022. El modelo estimado muestra resultados comparables en términos de precisión de pronóstico a los obtenidos con otras técnicas de Machine Learning como Ridge y Lasso, a la vez que su desempeño no es estadísticamente diferente al de la mediana de los pronosticadores profesionales relevados en el REM del BCRA. Hay mayor evidencia de un mejor desempeño respecto de los modelos de pronóstico tradicionales ARMA, ya que su Error Absoluto Medio es un desvío estándar menor al de los ARMA optimizados mes a mes mediante el criterio de Akaike.

Estos modelos constituyen entonces una herramienta complementaria de valor para el pronosticador, con la ventaja de que su estructura permite analizar no linealidades en el efecto de distintas variables respecto de su poder predictivo sobre la inflación. Asimismo, se trata de una metodología de relativa facilidad de estimación en comparación con modelos estructurales paramétricos o con los que requieren un mayor tratamiento de las variables involucradas para refinar sus impactos causales y/o limitar potenciales problemas de endogeneidad. Por este motivo, constituye una herramienta que aporta perspectivas con valor agregado respecto del inventario de modelos existentes, y simultáneamente requiere baja demanda de trabajo computacional para mejorar la precisión de los pronósticos de inflación de corto plazo.

Entre las limitaciones del modelo, la principal es que la necesidad de incorporar entornos de alta inflación tantos años atrás en el tiempo limita la cantidad de variables que pueden ser incorporadas, ya que la disponibilidad de series relativamente homogéneas en frecuencia mensual es más escasa. Fundamentalmente, sería relevante la inclusión de series mensuales que capten expectativas o consensos de mercado (aunque, como se dijo anteriormente, podrían estar parcialmente capturadas por el tipo de cambio paralelo y la inflación rezagada).

A pesar de que las series desde 1962 permiten una extensión de las series temporales por 744 períodos mensuales, esta longitud todavía constituye una muestra relativamente pequeña para realizar este tipo de estimaciones de Machine Learning. Por ello, los resultados presentan cierta volatilidad de acuerdo al muestreo aleatorio en cada iteración y es necesario estimar una varianza del pronóstico para tener una medida más precisa de la incertidumbre involucrada. Una mayor extensión temporal también permitiría incluir más indicadores sin perder valiosos grados de libertad en la estimación.

Por delante, la agenda de investigación y trabajo futuro es extensa. La literatura de pronósticos macroeconómicos recién en los últimos años ha comenzado a incorporar estas nuevas herramientas de big data y aprendizaje automático, por lo que hay mucho terreno inexplorado. En primer lugar, es posible hacer un refinamiento de este modelo computando Random Forest por cuantiles, de forma tal de no tener sólo estimaciones puntuales sino pronósticos de la distribución entera de probabilidades para la inflación de cada mes (tal como se hizo en Lenza et al, 2023, para la Eurozona). Además, evaluar otras metodologías como *Gradient Boosting* o redes neuronales podría aportar nuevas perspectivas y ganar precisión en la estimación de pronósticos, a pesar de que el requerimiento de datos para estos métodos es aún mayor y ello constituye un obstáculo difícil de sortear en Argentina. La incorporación de datos no estructurados como regresores (por ejemplo, fotos de portadas de diarios) podría ser otra hoja de ruta a explorar. Finalmente, también resta analizar la posibilidad de realizar combinaciones óptimas de pronósticos entre modelos para aumentar aún más la precisión alcanzada.

# Referencias Bibliográficas